\documentclass[aps,prd,twocolumn,tightenlines,showpacs,amsmath,amssymb,nofootinbib]{revtex4-2}

\usepackage{graphicx}
\usepackage{enumitem}						
\usepackage[colorlinks=true,linkcolor=.]{hyperref}
\usepackage{setspace}						
\usepackage[T1]{fontenc}					
\usepackage[utf8]{inputenc}					
\usepackage{booktabs}						
\usepackage[toc]{appendix}
\usepackage{amsmath, amssymb, amsfonts}			
\usepackage{mathtools}						
\usepackage{physics}						
\usepackage{bm} 							
\usepackage{slashed} 						
\usepackage{todonotes}   
\usepackage{hyperref}
\usepackage{acronym}

\newcommand*{\del}{\mathop{\mathrm{{}\partial}}\mathopen{}}

\begin{document}

\title{Operator spectrum of nonrelativistic CFTs at large charge}
\author{Vito Pellizzani}
\email{pellizzani@itp.unibe.ch}
\affiliation{Albert Einstein Center for Fundamental Physics, Institute for Theoretical Physics, University of Bern, Switzerland}
\date{\today}

\begin{abstract}
We extend and clarify the large-charge expansion of the conformal dimension $\Delta_Q$ of the lowest operator of charge $Q$ in nonrelativistic conformal field theories using the state-operator correspondence. The latter requires coupling the theory to an external harmonic trap that confines the particles to a spherical cloud, at the edge of which the effective theory breaks down and leads to divergences. Only recently has this issue been overcome by constructing appropriate counterterms at the edge of the cloud [\href{https://arxiv.org/abs/2010.07967}{arXiv:2010.07967}]. In this paper, we extend these results by systematically analyzing the degree of divergence of operators in the effective action and show that there always exist appropriate edge counterterms that make the final contributions to $\Delta_Q$ finite. On the other side of the correspondence, this also provides new corrections to the Thomas-Fermi approximation of the unitary Fermi gas, and we comment on their relevance for ultracold atom physics.
\end{abstract}

\maketitle


\acrodef{lo}[\textsc{lo}]{leading-order}
\acrodef{nlo}[\textsc{nlo}]{next-to-leading-order}
\acrodef{lg}[\textsc{lg}]{Landau--Ginzburg}
\acrodef{bcs}[\textsc{bcs}]{Bardeen-Cooper-Schrieffer}
\acrodef{bec}[\textsc{bec}]{Bose-Einstein condensate}
\acrodef{gp}[\textsc{gp}]{Gross-Pitaevskii}
\acrodef{ccwz}[\textsc{ccwz}]{Callan--Coleman--Wess--Zumino~\cite{Coleman:1969sm,Callan:1969sn}}
\acrodef{kk}[\textsc{kk}]{Kaluza--Klein}
\acrodef{sym}[\textsc{sym}]{super Yang--Mills}
\acrodef{vev}[\textsc{vev}]{vacuum expectation value}
\acrodef{cs}[\textsc{cs}]{Chern--Simons}
\acrodef{scft}[\textsc{scft}]{super-conformal field theory}
\acrodefplural{scft}[\textsc{scft}s]{super-conformal field theories}
\acrodef{qft}[\textsc{qft}]{quantum field theory}
\acrodefplural{qft}[\textsc{qft}s]{quantum field theories}
\acrodef{cft}[\textsc{cft}]{conformal field theory}
\acrodefplural{cft}[\textsc{cft}s]{nonrelativistic conformal field theories}
\acrodef{nrcft}[\textsc{nrcft}]{nonrelativistic conformal field theory}
\acrodefplural{nrcft}[\textsc{nrcft}s]{nonrelativistic conformal field theories}
\acrodef{eft}[\textsc{eft}]{effective field theory}
\acrodef{qcd}[\textsc{qcd}]{quantum chromodynamics}
\acrodef{ope}[\textsc{ope}]{operator product expansion}
\acrodef{dof}[\textsc{dof}]{degree of freedom}
\acrodefplural{dof}[\textsc{dof}]{degrees of freedom}
\acrodef{lsm}[\textsc{lsm}]{linear sigma model}
\acrodef{nlsm}[\textsc{nlsm}]{nonlinear sigma model}
\acrodef{ir}[\textsc{ir}]{infrared}
\acrodef{uv}[\textsc{uv}]{ultraviolet}
\acrodef{eom}[\textsc{eom}]{equation of motion}
\acrodefplural{eom}[\textsc{eom}]{equations of motion}
\acrodef{adscft}[\textsc{ads-cft}]{\textsc{ads-cft}}
\acrodef{sct}[\textsc{sct}]{special conformal transformation}
\acrodef{rg}[\textsc{rg}]{renormalization group}

\section{Introduction} \label{sec:introduction}

The large-charge approach to strongly coupled systems with global symmetries is a systematic way of deriving the spectrum of charged operators in an expansion in inverse powers of the charge, as first discussed in \cite{Hellerman:2015nra} (see \cite{Gaume:2020bmp} for a recent review), as well as certain correlation functions \cite{Monin:2016jmo,Cuomo:2021ygt,Cuomo:2020rgt}. Sequels of this approach include the large-charge expansion in \acp{nrcft} \cite{Hellerman:2020eff,Orlando:2020idm,schroedinger,Kravec:2018qnu,Kravec:2019djc}, the $O(N)$ model at large charge \cite{Alvarez-Gaume:2016vff}, its double-scaling large-$N$ limit \cite{Alvarez-Gaume:2019biu,Giombi:2020enj} and the study of nonperturbative corrections thereof using resurgence techniques \cite{Dondi:2021buw}, the large $R$-charge limit \cite{Hellerman:2017veg,Hellerman:2017sur,Bourget:2018obm,Hellerman:2018xpi,Beccaria:2018xxl,Beccaria:2020azj,Hellerman:2020sqj} and the $\epsilon$-expansion at large charge \cite{Badel:2019oxl,Arias-Tamargo:2019xld,Watanabe:2019pdh,Antipin:2020abu,Antipin:2020rdw,Antipin:2021akb}, among others. In most cases, the state-operator correspondence turns out to be extremely powerful.

In this paper, we are concerned with \acp{nrcft} where the correspondence maps the spectrum of conformal dimensions of (positively charged) local operators to the energy spectrum of states in an external spherical harmonic trap
\begin{equation}
	A_0(\vec x) = \frac{m \omega^2}{2 \hbar} |\vec x|^2,
\end{equation}
and vice-versa \cite{PhysRevA.74.053604,Nishida:2007pj,Goldberger:2014hca}. In particular, we focus on the conformal dimension $\Delta_Q$ of the lowest operator of fixed charge $Q \gg 1$, which can be accessed via the ground-state energy $E_0$ of the trapped system with $Q$ particles confined to a spherical cloud. The argument is presented for general spatial dimension $d$ in dimensionless units $\hbar = m = \omega = 1$, in which case we simply have $\Delta_Q = E_0$. 

In order to derive the large-charge expansion of the ground-state energy, we construct the \ac{eft} for the Goldstone boson $\chi$ associated with the broken $U(1)$ (i.e. particle number) symmetry with appropriate dilaton dressing rules that guarantee conformal invariance. To leading order, this description corresponds to the usual Thomas-Fermi approximation, and the first subleading corrections were found by Son and Wingate in a small momentum expansion for the \ac{nlsm} \cite{Son:2005rv}. While the power counting they used in this work might seem somewhat arbitrary, it is in fact best understood from a large-charge perspective \cite{Kravec:2018qnu} (see also \cite{schroedinger}). However, this effective theory is known to break down close to the edge of the cloud, where the particle density falls off and gives rise to divergences even at the classical level \cite{Son:2005rv, Kravec:2018qnu, Orlando:2020idm}. Building upon the recent work \cite{Hellerman:2020eff}, we classify all types of edge divergences and explain how to restore tree-level consistency by constructing appropriate counterterms in a procedure that we refer to as $\delta_\epsilon$-layer regularization. We show that, in general,
\begin{equation}
\begin{aligned}
	\Delta_Q & = Q^\frac{d+1}{d} \left[ a_1 + a_ 2 Q^{-\frac{2}{d}} + a_3 Q^{-\frac{4}{d}} + \ldots \right] \\
	& + Q^\frac{2d-1}{3d} \left[ b_1 + b_ 2 Q^{-\frac{2}{3d}} + b_3 Q^{-\frac{4}{3d}} + \ldots \right] \\
	& + Q^\frac{d-5}{3d} \left[ c_1 + c_ 2 Q^{-\frac{2}{3d}} + c_3 Q^{-\frac{4}{3d}} + \ldots \right] + \ldots
\end{aligned}
\end{equation}
plus quantum corrections starting at $Q^0$. The first line is analogous to the relativistic case, while the second and third lines are new structures arising from edge effects. Note that some of the $b_i$'s contain a $\log Q$-factor when $d$ is even, e.g.
\begin{equation} \label{eq:intro_2D_DeltaQ}
\begin{aligned}
	\Delta_Q^{(d=2)} = c_1 Q^\frac{3}{2} & + c_2 \sqrt{Q} \log Q + c_3 \sqrt{Q} \\
	& + c_4 Q^\frac{1}{6} - 0.29416 + \ldots,
\end{aligned}
\end{equation}
where the last term is the model-independent one-loop Casimir energy calculated in \cite{Orlando:2020idm}, and further corrections scale with negative powers of $Q$. Equation (\ref{eq:intro_2D_DeltaQ}) was first derived in \cite{Hellerman:2020eff}, based on \cite{Orlando:2020idm,schroedinger,Kravec:2018qnu}, although we shall clarify the origin of the $Q^\frac{1}{2}$ and $Q^\frac{1}{6}$ terms. Similarly, we show that
\begin{equation} \label{eq:intro_3D_DeltaQ}
\begin{aligned}
	\Delta_Q^{(d=3)} = c_1 Q^\frac{4}{3} & + c_2 Q^\frac{2}{3} + c_3 Q^\frac{5}{9} \\
	& + c_4 Q^\frac{1}{3} + c_5 Q^\frac{1}{9} + c_6 Q^0 + \ldots,
\end{aligned}
\end{equation}
and we argue that one should expect the presence of a universal $\log Q$-term associated with the Casimir energy, which will be computed in an upcoming publication \cite{Hellerman:2021qzz}. Son \& Wingate \cite{Son:2005rv} derived the first two terms of Eq.~(\ref{eq:intro_3D_DeltaQ}) and anticipated the presence of a divergent term that would scale like $Q^\frac{5}{9}$ after $\delta_\epsilon$-layer regularization. In this paper, we renormalize it for the first time and extend the expansion down to $Q^0$.

Moreover, the description of the system in terms of the dilaton mode allows us to explore the near-conformal regime of the theory by introducing a small dilaton mass $m_\sigma$ \emph{à la} Coleman \cite{Coleman:1988aos}, as was done in \cite{Orlando:2020idm} based on the general expectation detailed in \cite{PhysRevD.103.105026,PhysRevD.101.065018} that a dilatonlike mode appears near a smooth quantum phase transition. We overcome the issues faced in \cite{Orlando:2020idm} related to boundary divergences and find that the signature of this mass deformation is an additional $\sqrt Q \log Q$ and $\sqrt Q$-contributions in $d = 3$, while the structure of the expansion is unaffected in $d = 2$. Note that, while the concept of conformal dimension becomes (softly) ill-defined in this scenario, the corrections induced by the dilaton mass to the ground-state energy of the trapped system are interesting \emph{per se}. In this work, one can therefore think of $\Delta_Q$ as the latter energy, which exactly corresponds to the conformal dimension of the lowest operator of charge $Q$ only when $m_\sigma = 0$.



\medskip
Finally, let us comment on the relevance for ultracold atom physics, as trapped gases can be realized experimentally (see e.g. the beautiful reviews \cite{Dalfovo_1999,RevModPhys.80.1215} and references therein). Typically, this is achieved for cold and dilute atomic Fermi gases whose interaction strength is dominated by the s-wave scattering and can be tuned using Feshbach resonances. Correspondingly, the value and even the sign of the dimensionless scattering parameter $\frac{1}{k_F a_s}$---where $k_F$ is the Fermi wave-vector and $a_s$ the s-wave scattering length---can be changed. In the \ac{bcs} regime $\frac{1}{k_F a_s} \ll -1$, the interaction is weakly attractive and fermions form Cooper pairs, while for $\frac{1}{k_F a_s} \gg 1$, the attraction is strong and binds pairs of fermions with opposite spin together. The latter system is effectively described by a weakly interacting bosonic gas of such molecules (also called dimers), i.e. a \ac{bec} \cite{RevModPhys.80.1215}. Both regimes are known to exhibit superfluidity and no phase transition occurs in between, indicating a smooth crossover that preserves superfluidity for all values of $k_F a_s$. This is particularly relevant for the crossover region $\frac{1}{k_F a_s} \in [-1, 1]$, centered around the resonant case $\frac{1}{k_F a_s} = 0$ known as the \emph{unitary} limit, where the system is strongly interacting and an expansion in $k_F a_s$ is inappropriate. While a complete description of the crossover is still lacking, the emergent scale invariance at unitarity allows for an \ac{eft} description of the cold Fermi gas with a large number of trapped particles, as initiated by Son \& Wingate \cite{Son:2005rv} and completed in the present work from a \ac{lsm} perspective, where the only massless, low-energy degree of freedom $\chi$ corresponds to the phase of the condensate. As already mentioned, this \ac{eft} goes beyond the Thomas-Fermi approximation, yielding corrections e.g. to the ground-state energy, Eq.~(\ref{eq:intro_3D_DeltaQ}), or the doubly integrated density $n(x_3) = \iint \dd x_1 \dd x_2 \, \rho(\vec x)$ measured experimentally in \cite{PhysRevLett.92.120401} (where $\rho(\vec x)$ is the charge density):
\[
\begin{aligned}
	n(x_3) & = \frac{2 \pi g}{5} \left[ \frac{2 c_\frac{4}{3} \mu}{g} \left( 1 - \frac{x_3^2}{2 \mu} \right) \right]^\frac{5}{2} \\
	& \times \left[ 1 + \frac{45 c_\frac{2}{3}}{32 c_\frac{4}{3} \mu^2} \left\{ \frac{5}{\left( 1 - \frac{x_3^2}{2 \mu} \right)^2} - \frac{1}{\left( 1 - \frac{x_3^2}{2 \mu} \right)^3} \right\} + \ldots \right].
\end{aligned}
\]
Here, $\mu$ is the chemical potential and $g$, $c_\frac{4}{3}$, $c_\frac{2}{3}$ are Wilsonian parameters. The latter is associated with the simplest subleading operator in the \ac{eft}; without it, only the first line above matters, which can be written as $n(x_3) = \frac{16}{5 \pi} \frac{Q}{R_{cl}} \left( 1 - \frac{x_3^2}{R_{cl}^2} \right)^\frac{5}{2}$, where $R_{cl} = 2 \mu$ is the classical radius of the cloud, thus matching the known expression \cite{RevModPhys.80.1215}. Further corrections can readily be computed. It would also be interesting to investigate to what extent the somewhat naïve breaking of conformal invariance caused by the introduction of a small dilaton mass mentioned earlier allows for the exploration of the crossover region (e.g. in the spirit of \cite{Escobedo:2009bh}).

Another interesting direction for future research in ultracold atom physics concerns \ac{bec}, for which the Gross-Pitaevskii theory predicts that the ground-state energy be given by \cite{Dalfovo_1999}
\[
	E_0 = \int \dd^3 x \left[ \frac{1}{2} (\del_i a(\vec x))^2 + A_0(\vec x) a(\vec x)^2 + 2 \pi a_s \cdot a(\vec x)^4 \right],
\]
where $a(\vec x)$ is the radial mode of the condensate wave-function $\Phi(t, \vec x) = a(\vec x) e^{-i \mu t}$ and $\mu$ is the chemical potential. The first term in this expression is called \emph{quantum pressure} and is neglected in the Thomas-Fermi approximation. However, it becomes important close to the edge of the cloud, in a region sometimes called \emph{effective surface thickness} in this context. Upon approximating the potential by a linear ramp in this region, the authors of \cite{PhysRevA.58.3185} (based on \cite{PhysRevA.54.4213}) found schematically
\begin{equation}
	E_0 = d_1 Q^\frac{7}{5} + d_2 Q^\frac{3}{5} \log Q + d_3 Q ^\frac{3}{5} + \ldots,
\end{equation}
where we did not keep track of $a_s$ for simplicity. It would therefore be interesting to understand how much this result could be improved by including edge counterterms similar to the ones presented in the present work.

\begin{center} * * * \end{center}

This paper is organized as follows. In Sec.~\ref{sec:dilaton_dressing} and \ref{sec:LO_lagr}, we review the construction of the leading-order effective action from the \ac{lsm} perspective using the dilaton dressing, following \cite{Orlando:2020idm}. We then discuss in Sec.~\ref{sec:dressing_rules} how to adapt the previous dilaton dressing as we approach the edge of the cloud, as first discussed in \cite{Hellerman:2020eff}, and we give the recipe for the construction of edge counterterms in Sec.~\ref{sec:counterterms}, based on the same reference. The core of our work is presented in Sec.~\ref{sec:subleading}, where we analyze the possible diverging behaviors of operators due to boundary effects and show that the previously constructed counterterms always match. This allows to complete the large-charge expansion of $\Delta_Q$ in any dimension up to terms that scale with negative powers of $Q$, in which case quantum corrections have to be taken into account. We make some general observations and comment on quantum corrections in Sec.~\ref{sec:properties}, and finally work out the $d=2$ and $d=3$ cases in Sec.~\ref{sec:ex_2D} and Sec.~\ref{sec:ex_3D}, respectively, including the aforementioned dilaton mass deformation.

\section{Effective action} \label{sec:effective_action}
\subsection{Dilaton dressing and radial mode} \label{sec:dilaton_dressing}

Working in a sector of fixed charge spontaneously breaks the associated global symmetry, as well as conformal invariance. Describing the \ac{eft} in terms of the Goldstone mode that accounts for the breaking of conformal invariance, namely the dilaton $\sigma$, turns out to be convenient even though it may actually be massive (see e.g. \cite{Brauner:2014aha} for a discussion on gapped Goldstone bosons). A simple construction is due to Coleman \cite{Coleman:1988aos}, who pointed out that it is possible to promote Lorentz invariance of a given theory to a full conformal invariance by dressing the operators with an appropriate factor involving the dilaton. The very same game can be played with nonrelativistic theories \cite{Orlando:2020idm, Arav:2017plg}, where the nonrelativistic conformal symmetry group is usually referred to as the \emph{Schrödinger group}. Scale transformation $(t, \vec x) \to (e^{z\tau} t, e^\tau \vec x)$---where $z=1$ in the relativistic case and $z=2$ in the nonrelativistic one---acts on the dilaton as
\begin{equation}
	\sigma(t, \vec x) \longrightarrow \sigma(t, \vec x) + \frac{d + z - 2}{2 f},
\end{equation}
where the dimensionful parameter $f$ can be regarded as the (inverse) decay constant of the dilaton \cite{Komargodski:2011vj}.

If one considers a theory featuring a global $U(1)$ symmetry, as is the case of the Schrödinger group, one may first construct the most general \ac{eft} for the Goldstone mode $\chi$ invariant under Galilean or Lorentz symmetry that nonlinearly realizes the $U(1)$ symmetry, and then appropriately dress the operators with the dilaton $\sigma$ so as to make them marginal. Note that in general, these two fields can then be conveniently recast as
\begin{equation}
	\psi = \frac{1}{f} e^{-f \sigma - i \chi},
\end{equation}
and we shall therefore refer to $a \equiv |\psi| = \frac{1}{f} e^{-f \sigma}$ as the \emph{radial mode}. Of course, this construction would require an infinite number of Wilsonian coefficients, but one can then organize them in a large-charge expansion and truncate to any desired order. While this provides an explicit recipe for the construction of the large-charge \ac{lsm} of the effective theory (which essentially works in the same way for the relativistic \cite{Gaume:2020bmp,Orlando:2019skh} and the nonrelativistic cases \cite{Orlando:2020idm}), it should be pointed out that the radial mode becomes massive under spontaneous breaking of the $U(1)$ symmetry and thus decouples below the energy scale associated with the charge. Upon integrating it out, one would recover an equivalent large-charge effective action for the Goldstone $\chi$ alone in the form of a \ac{nlsm}, which can be obtained using different methods, e.g. the coset construction \cite{Monin:2016jmo, Kravec:2018qnu}. In this paper, we shall use the \ac{lsm} description in view of including a small dilaton mass deformation.

\subsection{Leading-order Lagrangian} \label{sec:LO_lagr}

As mentioned in the introduction, we aim to compute the conformal dimension of the lowest operator at large charge using the nonrelativistic state-operator correspondence. Accordingly, we consider the theory coupled to an external trapping potential,
\begin{equation}
		A_0(r) = \frac{1}{2} r^2,
\end{equation}
which restricts the support of the (classical) theory to a ball of finite radius, i.e., a {\em cloud}---or droplet---of particles at the edge of which the particle density rapidly falls off to zero. Unlike in the relativistic case where the state-operator correspondence is realized on a fixed background, the cloud is a dynamical object whose boundary undergoes quantum fluctuations. It is a very reasonable question to wonder whether short-distance physics causes any trouble close to the edge, and the answer is known to be positive \cite{Son:2005rv, Kravec:2018qnu, Orlando:2020idm}. This issue is already present at the classical level, and a sharp cutoff procedure was discussed in these references, where the so-called $\delta_\epsilon$-layer is removed at the edge in order to regularize the theory. More recently, a thorough discussion of the possible counterterms located at the edge of the droplet has been carried out in \cite{Hellerman:2020eff}. We aim at translating and extending these results into the language of the \ac{lsm} at large charge.

The building block of a generic Galilean invariant theory for the Goldstone mode $\chi$ in the trap is the operator
\begin{equation}
		U \equiv \dot\chi - A_0(r) - \frac{1}{2} (\del_i \chi)^2,
\end{equation}
where the presence of $A_0(r)$ requires some notion of general coordinate invariance, as discussed in \cite{Son:2005rv}. As we shall see later, derivatives of this operator, as well as other operators featuring more derivatives of the Goldstone mode $\chi$ contribute to the effective action but, for now, let us focus on the power series in $U$,
\begin{equation}
		\mathcal{L}(\chi) = -k_0 + \sum_{n=1}^\infty k_n U^n,
\end{equation}
where the $k_i$'s are Wilsonian coefficients. Promoting such a Galilean-invariant Lagrangian to a fully Schrödinger-invariant one is now an easy task with the dilaton dressing. Since the dimension of the radial mode $a = \frac{1}{f} e^{-f \sigma}$ is $[a] = \frac{d}{2}$ and $[U] = 2$, we simply have
\begin{equation} \label{eq:lagr_series}
		\mathcal{L}(\chi, a) = -k_0 a^{2+\frac{4}{d}} + a^{2+\frac{4}{d}} \sum_{n=1}^\infty k_n \cdot \left( \frac{U}{a^\frac{4}{d}} \right)^n.
\end{equation}
In the superfluid ground state, the \ac{vev} of the Goldstone mode is $\langle \chi \rangle = \mu \cdot t$, where $\mu$ is the chemical potential. The $U(1)$ and conformal symmetries are spontaneously broken, and the equation of motion for $a$ imposes that the ratio $\frac{U}{a^\frac{4}{d}}$ is necessarily a constant. Correspondingly, their \ac{vev} are of the form
\begin{equation} \label{eq:a_and_U}
	\langle a \rangle^\frac{4}{d} \sim \langle U \rangle = \mu \left( 1 - \frac{r^2}{2 \mu} \right).
\end{equation}
From the form of the action, Eq.~(\ref{eq:lagr_series}), one readily sees that $a$ acquires a mass $m_a^2 \sim \langle a \rangle^\frac{4}{d} \sim \mu$. Moreover, the ground-state charge density $\rho \propto \langle a \rangle^2$ is supported on the interval $r \in [0, R_{cl}]$ where $R_{cl} \equiv \sqrt{2 \mu}$ defines the radius of the cloud, i.e. the classical turning point, and sets an \ac{ir} length-scale. Upon integrating the charge density over this region, one finds that the total charge is
\begin{equation}
	Q \sim \mu^d.
\end{equation}
If one associates an \ac{uv} length-scale $R_\mu = \sqrt{\frac{2}{\mu}}$ with the mass of the radial mode, the effective theory description is under perturbative control when there is a separation of scales,
\begin{equation}
		R_{cl} \gg r \gg R_{\mu},
\end{equation}
which amounts to requiring that the controlling parameter $\frac{R_\mu}{R_{cl}} = \frac{1}{\mu} \sim Q^{-\frac{1}{d}}$ be small. This, in turn, is guaranteed by the large-charge condition $Q \gg 1$.

The drawback of keeping track of the massive mode $a$ in the low-energy description is that we technically have to account for series of operators, as in Eq.~(\ref{eq:lagr_series}), that give the same contribution to observables to leading-order. Roughly speaking, integrating the radial mode out in Eq.~(\ref{eq:lagr_series}) gives a single leading-order term $U^{1+\frac{2}{d}}$ in the \ac{nlsm}, and trading $a^\frac{4}{d}$ for $U$ is therefore unseen at the level of the \ac{nlsm}. The minimal Lagrangian that captures all the above properties is given by
\begin{equation} \label{eq:lagr_LO}
	\mathcal{L}_{LO}(\chi, a) = c_\frac{d+1}{d} a^2 U - \frac{d}{2 (d + 2)} g a^{2+\frac{4}{d}},
\end{equation}
where we renamed and rescaled the Wilsonian coefficients for future convenience. Correspondingly, the ground-state energy---and therefore, the conformal dimension of the lowest operator of charge $Q$ in the system without trap---is given by
\begin{equation} \label{eq:Delta_LO}
	\Delta_Q = \frac{d}{d + 1} \zeta Q^\frac{d+1}{d},
\end{equation}
where $\zeta = \sqrt{\frac{g}{4 \pi c_\frac{d+1}{d}}} \left[ \frac{2 \Gamma(d)}{c_\frac{d+1}{d} \Gamma\left(\frac{d}{2}\right)} \right]^\frac{1}{d}$ is a constant, in accordance with the \ac{nlsm} results \cite{Kravec:2018qnu}. The advantage of this description, however, is that it allows for a rather straightforward analysis of the subleading corrections to Eq.~(\ref{eq:Delta_LO}), as discussed in Sec.~\ref{sec:subleading}.

In order to further simplify the argument, we introduce the dimensionless coordinate $z \equiv 1 - \frac{r^2}{R_{cl}^2} = 1 - \frac{r^2}{2 \mu}$, which measures the distance from the classical boundary of the cloud. Since spherical symmetry is preserved by the superfluid ground state, it will be convenient to express every \ac{vev} as a function of $z$. Useful properties are
\begin{equation} \label{eq:z_stuff}
\begin{aligned}
	& (\del_i f(\vec x)) (\del_i g(\vec x)) = \frac{2(1-z)}{\mu} f'(z) g'(z), \\
	& \nabla^2 f(\vec x) = \frac{2}{\mu} \left[ (1-z) f''(z) - \frac{d}{2} f'(z) \right], \\
	& \int_{cloud} \dd^d x \, f(\vec x) = \frac{(2 \pi \mu)^\frac{d}{2}}{\Gamma\left(\frac{d}{2}\right)} \int_0^1 \dd z \, (1 - z)^\frac{d-2}{2} f(z),
\end{aligned}
\end{equation}
where primes refer to derivatives with respect to $z$ and $f, g$ are spherically invariant functions. Note that spatial derivatives of operators \emph{a priori} make their contributions to the conformal dimension $\Delta_Q$ parametrically smaller due to the division by $\mu \sim Q^\frac{1}{d}$.

At this stage, let us point out that Eq.~(\ref{eq:lagr_LO}) in $d = 3$ corresponds to the Thomas-Fermi approximation of the unitary Fermi gas, and yields, among others, the known expression for the doubly integrated density mentioned in the introduction and measured experimentally in \cite{PhysRevLett.92.120401}, namely
\begin{equation} \label{eq:profile_LO}
\begin{aligned}
	n(x_3) & = \iint \dd x_1 \dd x_2 \, \rho(\vec x) \\
	& = \frac{2 \pi g}{5} \left[ \frac{2 c_\frac{4}{3} \mu}{g} \left( 1 - \frac{x_3^2}{2 \mu} \right) \right]^\frac{5}{2},
\end{aligned}
\end{equation}
where the $x_i$'s take values in the cloud. At the end of this paper, we discuss corrections to this expression.

\subsection{Dressing rules} \label{sec:dressing_rules}

The presence of the dilaton field, via the radial mode $a(t, \vec x) = \frac{1}{f} e^{-f \sigma(t, \vec x)}$, allows for the dressing of operators to marginality, as discussed in the previous section. At the same time, the breakdown of the effective theory near the edge of the cloud is associated with the vanishing of the particle density and, hence, the vanishing of $a$ \cite{Son:2005rv, Kravec:2018qnu, Orlando:2020idm}. From a large-charge perspective, this indicates that the dressing rule based on powers of $a$ is only appropriate when edge effects are negligible, i.e. in the \emph{bulk} of the cloud (to be defined later). As we approach the boundary, the dressed theory fails to describe the system, and another nonvanishing, nonsingular operator needs to take over as the new appropriate dressing rule \cite{Hellerman:2020eff}.

Concretely, a generic dressing operator can involve powers of $a$ and its derivatives $(\del_i a)^2$, so we consider
\begin{equation}
	\mathcal{D}_{b,c} \equiv \left[ a^{2b} (\del_i a)^{2c} \right]^\frac{2}{d \cdot (b + c) + 2 c},
\end{equation}
where $b, c$ can be any positive numbers for now, and the overall power is chosen such that its dimension is fixed:
\begin{equation}
	[\mathcal{D}_{b,c}] = 2.
\end{equation}
Indeed, note that $[a^2] = d$ and $[(\del_i a)^2] = d+2$. It is straightforward to see that the dressing rule associated with $\mathcal{D}_{b,c}$ for an operator $\mathcal{O}$ with dimension $[\mathcal{O}]$ is
\begin{equation}
	\mathcal{O}_{dressed} \equiv \mathcal{O} \cdot \mathcal{D}_{b,c}^{\frac{d + 2 - [\mathcal{O}]}{2}}.
\end{equation}
 So how do we fix $b$ and $c$ in the bulk and at the edge? Since the dressing operator has the same dimension for any pair $(b, c)$, the selection criteria are rather simple \cite{Hellerman:2020eff}. To leading-order in $\mu$, the \ac{vev} of the dressing operator scales like
\begin{equation} \label{eq:mu_Dab}
	\langle \mathcal{D}_{b,c} \rangle \sim \mu^{1 - \frac{4 c}{d \cdot (b + c) + 2 c}} \cdot z^{1 - \frac{6 c}{d \cdot (b + c) + 2 c}},
\end{equation}
since $\langle a \rangle \sim (\mu \cdot z)^\frac{d}{4}$ and $(\del_i \langle a \rangle)^2 \sim \mu^\frac{d-2}{2} \cdot z^\frac{d-4}{2}$. In the bulk of the cloud, $z$ is of order 1 and the dressing rule is associated with the operator $\mathcal{D}_{b,c}$ that has the highest $\mu$-scaling, i.e. $c=0$. This yields the natural dressing rule used in the previous section, namely
\begin{equation} \label{eq:dress_bulk}
	\mathcal{D}_{bulk} \equiv a^\frac{4}{d}.
\end{equation}

At the edge, however, the dressing operator is required to be nonvanishing (unlike $\mathcal{D}_{bulk}$), and nonsingular. In short, its leading-order dependence on $\mu$ in the ground state should feature neither positive nor negative powers of $z$, i.e. it is a constant. This happens when $d \cdot (b + c) = 4 c$, and we get
\begin{equation} \label{eq:dress_edge}
	\mathcal{D}_{edge} \equiv \left[ a^{\frac{8}{d}-2} (\del_i a)^2 \right]^\frac{1}{3}.
\end{equation}
This operator is proportional to $\left| \del_i \left(a^\frac{4}{d}\right) \right|^\frac{2}{3}$ which is equivalent to the edge dressing rule originally discussed in \cite{Hellerman:2020eff} upon trading $a^\frac{4}{d}$ for $U$ at the level of the \ac{nlsm}.

\subsection{$\delta_\epsilon$-layer and edge counterterms} \label{sec:counterterms}

In order to account for the lack of control at the droplet edge, one can effectively cut off a small layer close the classical boundary of the cloud, as discussed in \cite{Son:2005rv, Kravec:2018qnu, Orlando:2020idm}. Following \cite{Hellerman:2020eff}, this regularization prescription can be made slightly more precise in terms of the dressing operators we have just found. Indeed, we eventually want to renormalize the theory by introducing counterterms in the region where $\mathcal{D}_{bulk} \sim \mathcal{D}_{edge}$ and beyond which $\mathcal{D}_{edge}$ is the only appropriate dressing operator. This condition is equivalent to
\begin{equation} \label{eq:regul_condition}
	(\del_i a)^2 \sim a^{\frac{4}{d}+2},
\end{equation}
which, in the ground state, is satisfied when $z \sim \mu^{-\frac{2}{3}}$, and we thus define the $\delta_\epsilon$-layer,
\begin{equation}
		\delta_\epsilon \equiv \frac{\epsilon}{\mu^\frac{2}{3}} \sim \mathcal{O}\left( Q^{-\frac{2}{3d}} \right),
\end{equation}
for an arbitrary constant $\epsilon \sim \mathcal{O}(1)$. We thereby define the bulk of the cloud as the region covered by the interval $\delta_\epsilon \lesssim z \leq 1$, in agreement with the literature.

In order to construct a counterterm at the edge, we use Eq.~(\ref{eq:dress_edge}) to dress to marginality an operator $\mathcal{O}$ of dimension $[\mathcal{O}]$ together with an operator-valued Dirac $\delta$-function $\delta(\mathcal{D}_{bulk})$ of dimension $-2$ \cite{Hellerman:2020eff}:
\begin{equation} \label{eq:edge_dressing}
\begin{aligned}
	\mathcal{O}_{edge} & \equiv \mathcal{O} \cdot \delta(\mathcal{D}_{bulk}) \cdot \mathcal{D}_{edge}^\frac{d+4-[\mathcal{O}]}{2} \\
	& = \mathcal{O} \cdot \delta(a^\frac{4}{d}) \cdot \left[ a^{\frac{8}{d}-2} (\del_i a)^2 \right]^\frac{d+4-[\mathcal{O}]}{6}.
\end{aligned}
\end{equation}
Given that the \ac{vev} $\langle a \rangle \equiv v(z)$ of the radial mode is of the form [cf. Eq.~(\ref{eq:a_and_U})]
\begin{equation}
	v(z) = v_{hom} \cdot z^\frac{d}{4} \cdot [1 + \mathrm{subleading}],
\end{equation}
where $v_{hom}$ is the superfluid ground-state solution of Eq.~(\ref{eq:lagr_LO}) in the homogeneous case (i.e. without trap), the Dirac $\delta$-function becomes
\begin{equation}
	\delta(v^\frac{4}{d}) = \delta\left( \frac{v^\frac{4}{d}}{v_{hom}^\frac{4}{d}} \right) \frac{1}{v_{hom}^\frac{4}{d}} = \frac{\delta(z)}{v_{hom}^\frac{4}{d}}.
\end{equation}
This shows that, in the ground state, counterterms are indeed located at $z = 0$, that is, at the classical edge.

The renormalization procedure thus consists in (1) regularizing the divergent integrals of operators dressed in the bulk by removing the $\delta_\epsilon$-layer from the domain of integration, and (2) introducing edge counterterms whose coefficients get renormalized so as to absorb the resulting $\epsilon$-dependence. After the first step, the regulator $\epsilon$ appears in logarithms and denominators, and thus serves as a diagnosis of divergences. However, the limit $\epsilon \to 0$ is not implicitly understood at any point, since no physical quantity depends on it.

At this stage, it is worth mentioning that edge counterterms have already been discussed in the literature in the context of effective open strings \cite{Hellerman:2016hnf} and, while this construction is very clear from an effective point of view, it would be interesting to put it on a more formal basis.

\subsection{Subleading operators} \label{sec:subleading}

We now make a simple argument that allows to push the large-charge expansion of $\Delta_Q$ beyond all known results so far. From a Wilsonian perspective, the effective action contains infinitely many operators. However, in a large-charge regime, there is a way in which we can organize them. The action Eq.~(\ref{eq:lagr_LO}) provides the leading-order contribution to the conformal dimension $\Delta_Q$, Eq.~(\ref{eq:Delta_LO}), which is of order $Q^\frac{d+1}{d}$. Operators with more derivatives yield corrections to $\Delta_Q$ that are parametrically suppressed by inverse powers of the charge. If one truncates the large-charge expansion of $\Delta_Q$ to a desired order in $Q$, the task is to identify the operators that survive in the action. In turn, this means we have to understand how a generic operator contributes to $\Delta_Q$.

To do so, let us consider an operator $\mathcal{O}$ with dimension $[\mathcal{O}]$, which we dress to marginality in the bulk as
\begin{equation} \label{eq:bulk_dressing}
	\mathcal{O}_{bulk} \equiv \mathcal{O} \cdot a^{\frac{2}{d}(d+2-[\mathcal{O}])}.
\end{equation}
If the latter appears in the Hamiltonian density, its contribution to $\Delta_Q$ is then obtained by integrating its \ac{vev} over the volume of the cloud. Spherical invariance being preserved by the superfluid ground state, this computation simplifies if the \ac{vev} is expressed as a function of the dimensionless coordinate $z$, in which case it turns into an integration over $z \in [0, 1]$---as indicated in Eq.~(\ref{eq:z_stuff})---which may need to be regularized upon removing the $\delta_\epsilon$-layer. More specifically, let $\mu[\mathcal{O}]$ and $z[\mathcal{O}]$ be such that the \ac{vev} of the operator $\mathcal{O}$ to leading-order in $\mu$ takes the form
\begin{equation}
	\langle\mathcal{O}\rangle \sim \mu^{\mu[\mathcal{O}]} \cdot z^{z[\mathcal{O}]} + \mathrm{(subleading)}.
\end{equation}
The dressed operator in the bulk, Eq.~(\ref{eq:bulk_dressing}), then has a \ac{vev} that scales to leading order in $\mu$ as
\begin{equation}
	\langle \mathcal{O}_{bulk} \rangle \sim \mu^{\mu[\mathcal{O}]+\frac{d+2-[\mathcal{O}]}{2}} \cdot z^{z[\mathcal{O}] + \frac{d+2-[\mathcal{O}]}{2}}.
\end{equation}
It is now straightforward to analyze the leading contribution of this operator to $\Delta_Q$. Indeed, if
\begin{equation}
	z[\mathcal{O}_{bulk}] \equiv z[\mathcal{O}] + \frac{d+2-[\mathcal{O}]}{2} \leq -1,
\end{equation}
a divergence occurs when integrating over $z \in [0, 1]$. In particular, a logarithmic divergence appears if $z[\mathcal{O}_{bulk}] = -1$. We thus regularize these divergences by removing the $\delta_\epsilon$-layer and, accounting for the factor of $\mu^\frac{d}{2} \sim \sqrt{Q}$ from the measure [cf. Eq.~(\ref{eq:z_stuff})], we find
\begin{widetext}
\begin{equation} \label{eq:DeltaQ_bulk_gen}
	\Delta_Q \ni
	\begin{cases}
		Q^{\frac{d+1}{d}-\frac{[\mathcal{O}]-2\mu[\mathcal{O}]}{2d}} & \text{if} \quad [\mathcal{O}] < d +4 + 2 z[\mathcal{O}] \\
		Q^{\frac{d+1}{d}-\frac{[\mathcal{O}]-2\mu[\mathcal{O}]}{2d}} \cdot \log \frac{Q}{\epsilon^{3d/2}} & \text{if} \quad [\mathcal{O}] = d +4 + 2 z[\mathcal{O}] \\
		\frac{Q^{\frac{2}{3}-([\mathcal{O}]+4z[\mathcal{O}]-6\mu[\mathcal{O}]-2)/(6d)}}{\epsilon^{([\mathcal{O}]-2z[\mathcal{O}]-d-4)/2}} & \text{if} \quad [\mathcal{O}] > d +4 + 2 z[\mathcal{O}].
	\end{cases}
\end{equation}
\end{widetext}
The last two cases are $\epsilon$-dependent and thus need to be renormalized using edge counterterms.

We address this issue in the following, but let us first answer a natural question: now that we have identified all possible (classical) leading contributions to the conformal dimension $\Delta_Q$, what sort of operator can $\mathcal{O}$ actually be? As dictated by general coordinate invariance \cite{Son:2005rv}, one possibility is given by
\begin{equation}
	Z \equiv \nabla^2 A_0 - \frac{1}{d^2} \left( \nabla^2 \chi \right)^2,
\end{equation}
whose \ac{vev} is $\langle Z \rangle = d$, but any other operator with more derivatives of $\chi$ actually has a vanishing \ac{vev}. Therefore, $\mathcal{O}$ is a composite operator made out of integer powers of $U$, $(\del_i U)^2$, $a$, $(\del_i a)^2$ and $Z$. As already mentioned, the massive radial mode $a$ would have to be integrated out and one can thus effectively trade $U$ for $a^\frac{4}{d}$---and likewise for $\del_i U$ and $\del_i a$---without changing the low-energy description. This is convenient because if $U$ only appears in the Lagrangian density as given in Eq.~(\ref{eq:lagr_LO}), then there is no need to worry about Legendre transforming $\mathcal{O}$: it simply enters the Hamiltonian density $\mathcal{H} = \frac{\del \mathcal{L}}{\del U} \dot\chi - \mathcal{L}$ with the opposite sign. Finally, we can strip off powers of $a$ as they will be restored appropriately upon dressing $\mathcal{O}$ in the bulk [cf. Eq.~(\ref{eq:bulk_dressing})]. Hence, we only need to consider operators of the form
\begin{equation} \label{eq:Omn}
		\mathcal{O}^{(m,n)} \equiv (\del_i a)^{2m} Z^n,
\end{equation}
with $m, n$ two positive integers and $[\mathcal{O}^{(m,n)}] = (d+2)m + 4n$, $\mu[\mathcal{O}^{(m,n)}] = \frac{d-2}{2} m$, $z[\mathcal{O}^{(m,n)}] = \frac{d-4}{2} m$.
The corresponding bulk operator is then given by
\begin{equation} \label{eq:Omn_bulk}
\begin{aligned}
	\mathcal{O}^{(m,n)}_{bulk} & \equiv (\del_i a)^{2m} Z^n \cdot a^{\frac{2}{d}((d+2)(1-m)-4n)} \\
	& = \left( \frac{(\del_i a)^2}{a^{\frac{4}{d}+2}} \right)^m \left( \frac{Z}{a^\frac{8}{d}} \right)^n a^{\frac{4}{d}+2},
\end{aligned}
\end{equation}
and its leading contribution to $\Delta_Q$ is classified as follows:
\begin{equation} \label{eq:DeltaQ_bulk}
	\Delta_Q \ni
	\begin{cases}
		Q^\frac{d+1-2(m+n)}{d} & \text{if} \quad 6m + 4n < d +4 \\
		Q^{\frac{2 d - 1}{3 d}-\frac{2n}{3d}} \cdot \log \frac{Q}{\epsilon^{3d/2}} & \text{if} \quad 6m + 4n = d +4 \\
		\frac{Q^{\frac{2 d - 1}{3 d}-\frac{2n}{3d}}}{\epsilon^{\frac{1}{2} (6m+4n-d-4)}} & \text{if} \quad 6m + 4n > d +4.
	\end{cases}
\end{equation}
Notice that there are infinitely many operators contributing to the same power of $Q$ in the last category, as it is independent of $m$. This remains true at the level of the \ac{nlsm}. For instance, the set of operators $(\del_i a)^{2 m}$ with $m \geq 2$ in $d = 2$ all give $Q^\frac{1}{2}$-contributions after $\delta_\epsilon$-regularization, but they have not yet appeared in the literature so far. The same holds for $(\del_i a)^{2 m} Z$ with $m \geq 1$ in $d = 2$, yielding $Q^\frac{1}{6}$-contributions, and similarly in $d = 3$ where sets of equally contributing operators give terms of order $Q^\frac{5}{9}$, $Q^\frac{1}{3}$, $Q^\frac{1}{9}$, etc. In Sec. \ref{sec:examples}, we illustrate this by including one operator of each set.

Note that these terms, however, appear with different powers of $\epsilon$ and can thus be distinguished and compensated for by a single edge counterterm. In order to identify the latter, we repeat the same analysis as before and remark that the only candidates to be dressed at the edge are of the form $Z^n$ (with $n$ a positive integer), since powers of $a$ and $(\del_i a)^2$ are appropriately incorporated by the dressing rule Eq.~(\ref{eq:edge_dressing}), which reads in this case
\begin{equation} \label{eq:Zn_edge}
	Z^n_{edge} \equiv Z^n \cdot \delta(a^\frac{4}{d}) \cdot \left[ a^{\frac{8}{d}-2} (\del_i a)^2 \right]^\frac{d+4(1-n)}{6}.
\end{equation}
After integrating the \ac{vev} of the latter over the volume of the cloud, the contribution to $\Delta_Q$ turns out to be
\begin{equation}
		\Delta_Q \ni Q^{\frac{2 d-1}{3d}-\frac{2n}{3d}},
\end{equation}
which matches exactly the regulator-dependent part (i.e. the last two categories) of Eq.~(\ref{eq:DeltaQ_bulk}). This analysis of divergences and counterterms thus provides us with a simple way of constructing the effective action that describes $\Delta_Q$ to a given order in $Q$. In Sec.~\ref{sec:examples}, we carry out this derivation in $d=2$ and $d=3$ up to corrections that scale with negative powers of the charge, but we first make some general observations.

\subsection{Properties} \label{sec:properties}

\emph{Equation of motion} (\textsc{eom})\acused{eom}. Consider the leading-order Lagrangian Eq.~(\ref{eq:lagr_LO}). The \ac{eom} with respect to the radial mode $a$ then simply reads $(v(z)/v_{hom})^\frac{4}{d} = z$ in the superfluid ground state, where $v_{hom} \equiv (2 \mu c_\frac{d+1}{d}/g)^\frac{d}{4}$ is the ground-state solution in the homogeneous case, i.e. when the trap is turned off. As we complement the Lagrangian with operators of the form Eq.~(\ref{eq:Omn_bulk}) and the corresponding counterterms, Eq.~(\ref{eq:Zn_edge}), the \ac{eom} gets more and more complicated, although it can always be put in the form
\begin{equation} \label{eq:eom_d}
	\left( \frac{v(z)}{v_{hom}} \right)^\frac{4}{d} = z [1 + B(z, v, v', v'')]
\end{equation}
in the ground state. In the bulk, $B(z, v, v', v'') \ll 1$ and one can solve this equation order by order in an expansion in $\frac{1}{\mu}$. For instance, adding the first subleading operator $-\frac{c_1}{2} \mathcal{O}^{(1,0)}_{bulk}$ to the Lagrangian yields
\begin{equation} \label{eq:v_NLO}
\begin{aligned}
	\left( \frac{v(z)}{v_{hom}} \right)^\frac{4}{d} & = z \left[ 1 - \frac{d}{16} \frac{c_1}{c_\frac{d+1}{d}} \frac{(4 - d) + (3 d - 4) z}{\mu^2 z^3} \right. \\
	& \hspace{11mm} \left. + \mathcal{O}\left( \frac{1}{\mu^4} \right) \right].
\end{aligned}
\end{equation}

\emph{Chemical potential.} The chemical potential can then be expressed as a function of the charge $Q$ by inverting
\begin{equation}
	Q = \int_{cloud} \dd^d x \, \rho(\vec x),
\end{equation}
where $\rho = c_\frac{d+1}{d} v(z)^2$ is the ground-state charge density. Using Eq.~(\ref{eq:v_NLO}) and removing the $\delta_\epsilon$-layer to regularize the divergent part, we find
\begin{equation}
	Q = \left( \frac{\mu}{\zeta} \right)^d \left[ 1 + \mathcal{O}\left( \mu^{-\frac{d+2}{3}} \right) \right],
\end{equation}
where
\begin{equation}
	\zeta \equiv \sqrt{\frac{g}{4 \pi c_\frac{d+1}{d}}} \left[ \frac{2 \Gamma(d)}{c_\frac{d+1}{d} \Gamma\left( \frac{d}{2} \right)} \right]^\frac{1}{d},
\end{equation}
as first discussed in \cite{Orlando:2020idm}. Therefore,
\begin{equation} \label{eq:mu_generic}
	\mu = \zeta Q^\frac{1}{d} \left[ 1 + \mathcal{O}\left( Q^{-\frac{d+2}{3d}} \right) \right].
\end{equation}
It is tempting to give an explicit expression for the correction in the square bracket based on the solution found above for $v(z)$, but some remarks are in order. Pushing the expansion further in Eq.~(\ref{eq:v_NLO}), one would actually face terms of the form $\frac{c_1^k}{(\mu^2 z^3)^k}$ ($k \in \mathbb{N}$), which all become of order one close to the edge (i.e. where $z \approx \delta_\epsilon \sim \mu^{-\frac{2}{3}}$), and contribute to the $Q^{-\frac{d+2}{3d}}$-correction in the chemical potential, which makes it hard to express it in closed form. A reasonable choice, though, is to limit ourselves from now on to linear order in the Wilsonian coefficients of subleading operators, since this does not change the nature of the expansion, but merely its coefficients. Similarly, the operator $-\frac{c_2}{4} \mathcal{O}^{(2,0)}_{bulk}$ also ends up contributing to the next-to-leading order in the chemical potential for exactly the same reason, and so does any operator $\mathcal{O}^{(m,0)}_{bulk}$, although we will not need to consider $m > 2$. With this,
\begin{equation}
\begin{aligned}
	\mu & = \zeta Q^\frac{1}{d} \left[ 1 + \frac{d^2 \Gamma(d)}{8 c_\frac{d+1}{d} \Gamma\left(\frac{d}{2}\right)^2} \frac{1}{Q^\frac{d+2}{3d}} \left\{ \frac{c_1}{\epsilon^\frac{4-d}{2}} \right. \right. \\
	& \hspace{16mm} \left. \left. + \frac{3 d^2}{32} \frac{c_2 g}{c_\frac{d+1}{d}} \frac{1}{\epsilon^\frac{10-d}{2}} \right\} + \mathcal{O}\left( Q^{-\frac{d+4}{3d}} \right) \right].
\end{aligned}
\end{equation}
Counterterms contributions would allow us to renormalize this expression and get rid of the $\epsilon$-dependence, but for practical purposes, we shall do this and fix the renormalized coefficients such that they cancel all divergences only at the very end of the computation of $\Delta_Q$. This is because the latter coefficients are only fixed up to a finite piece which we will not keep track of. Renormalizing $\mu$ at this stage would therefore just clutter the computations.

\emph{Structure of the expansion}. We now elaborate on the structure of the expansion of $\Delta_Q$ by first noting that the contribution of $\mathcal{O}^{(m,n)}_{bulk}$ is itself an expansion. Indeed, using the leading-order solution $v(z) \sim (\mu \cdot z)^\frac{d}{4}$ (corrections do not change the argument) and Eq.~(\ref{eq:z_stuff}), we have
\begin{equation}
\begin{aligned}
	\int_{cloud} \dd^d x \, \langle \mathcal{O}^{(m,n)}_{bulk} \rangle & \sim \mu^{d+1-2(m+n)} \int_0^1 \dd z \, \frac{(1 - z)^{\frac{d}{2}-1+m}}{z^{3m+2n-1-\frac{d}{2}}}.
\end{aligned}
\end{equation}
The integral on the right-hand side either converges and corresponds to the first case of Eq.~(\ref{eq:DeltaQ_bulk}), or needs to be regularized by setting the lower bound to $\delta_\epsilon$, yielding the last two cases of this classification. In the latter situation, however, the upper bound of the integral always gives a finite result, thus continuing the expansion in $Q^{-\frac{2}{d}}$ starting at $Q^\frac{d+1}{d}$. For concreteness, consider
\begin{equation}
	\mathcal{O}^{(1,1)}_{bulk} = \frac{(\del_i a)^2 Z}{a^\frac{8}{3}}
\end{equation}
in $d = 3$. We obtain
\[
\begin{aligned}
	\int_{cloud} \dd^3 x \, \langle \mathcal{O}^{(1,1)}_{bulk} \rangle & \sim \int_{\delta_\epsilon}^1 \dd z \, \frac{(1 - z)^\frac{3}{2}}{z^\frac{5}{2}} \\
	& = \pi + \frac{2}{3 \delta_\epsilon^\frac{3}{2}} - \frac{3}{\sqrt{\delta_\epsilon}} + \mathcal{O}\left( \sqrt{\delta_\epsilon} \right) \\
	& = \frac{2 \zeta}{3 \epsilon^\frac{3}{2}} Q^\frac{1}{3} - \frac{3 \zeta^\frac{1}{3}}{\sqrt{\epsilon}} Q^\frac{1}{9} + \pi + \mathcal{O}\left( Q^{-\frac{1}{9}} \right),
\end{aligned}
\]
where $\pi$ comes from the upper bound, and the rest is an expansion in $\delta_\epsilon$ whose dependence on $\epsilon$ is cured by the counterterms. All in all, the expansion of $\Delta_Q$ reads
\begin{equation} \label{eq:DeltaQ_result}
\begin{aligned}
	\Delta_Q & = Q^\frac{d+1}{d} \left[ a_1 + a_ 2 Q^{-\frac{2}{d}} + a_3 Q^{-\frac{4}{d}} + \ldots \right] \\
	& + Q^\frac{2d-1}{3d} \left[ b_1 + b_ 2 Q^{-\frac{2}{3d}} + b_3 Q^{-\frac{4}{3d}} + \ldots \right] \\
	& + Q^\frac{d-5}{3d} \left[ c_1 + c_ 2 Q^{-\frac{2}{3d}} + c_3 Q^{-\frac{4}{3d}} + \ldots \right] + \ldots
\end{aligned}
\end{equation}
The first line is completely analogous to the relativistic case, while the rest is specific to \ac{nrcft}. The last line arises when $\mu$ is replaced by $Q$, according to Eq.~(\ref{eq:mu_generic}). When $d$ is even, it can be absorbed in the second line, where some of the $b_i$'s contain $\log Q$-terms [see Eq.~(\ref{eq:DeltaQ_bulk})].

\emph{Casimir energy}. The leading quantum correction is due to the one-loop Casimir energy \cite{Hellerman:2020eff}, given by the Coleman-Weinberg formula applied to the spectrum of excited states, Eq.~(\ref{eq:spectrum_excit}) below. It is model-independent and \emph{a priori} enters the expansion of $\Delta_Q$ at order $Q^0$. However, it was shown in the relativistic case that the Casimir energy in odd spatial dimensions is divergent and yields instead a universal $Q^0 \log Q$-term after renormalization \cite{Cuomo:2020rgt}. This is hinted at by the presence of a classical $Q^0$-contribution that serves as a counterterm for this divergence, and the same phenomenon is thus to be expected in \ac{nrcft} when $d$ is odd. This is the object of a future publication \cite{Hellerman:2021qzz}.

\emph{Dilaton mass}. In addition to the operators discussed in the previous section, we now include a small dilaton mass deformation, as originally proposed by Coleman in \cite{Coleman:1988aos} in the relativistic case and studied in the context of \acp{nrcft} at large charge in \cite{Orlando:2020idm}. In the latter case, this potential is of the form
\begin{equation} \label{eq:Uc}
	U_C \equiv \left( \frac{d}{d + 2} \right)^2 \frac{m_\sigma^2}{4 f^2} \left[ (f a)^{2\frac{d+2}{d}} - 2 \frac{d + 2}{d} \log (f a) - 1 \right],
\end{equation}
where $m_\sigma \ll f^{-\frac{2}{d}}$ is a small dilaton mass, as can be seen from the fact that, to quadratic order, $U_C \approx \frac{1}{2} m_\sigma^2 \sigma^2$. Adding this term to the Lagrangian softly breaks conformal invariance and should trigger some signature in the ground-state energy of the trapped system.

\emph{Nonlinear sigma model}. If one is not interested in such a deformation, it might be more natural to work with the \ac{nlsm}, where the radial mode is integrated out. Upon trading $a^\frac{4}{d}$ for $U$, the bulk operators, Eq.~(\ref{eq:Omn_bulk}), become
\begin{equation}
\begin{aligned}
	\mathcal{O}^{(m,n)}_{bulk} & \equiv (\del_i U)^{2m} Z^n \cdot U^{\frac{d}{2}+1-(3m+2n)} \\
	& = \left( \frac{(\del_i U)^2}{U^3} \right)^m \left( \frac{Z}{U^2} \right)^n U^{\frac{d}{2}+1},
\end{aligned}
\end{equation}
and edge counterterms, Eq.~(\ref{eq:Zn_edge}), read
\begin{equation}
	Z^n_{edge} \equiv Z^n \cdot \delta(U) \cdot (\del_i U)^\frac{d+4(1-n)}{3}.
\end{equation}

\emph{Collective excitations}. As a final comment, let us mention that the energy spectrum of collective excitations above the ground state (see e.g. \cite{PhysRevA.62.041601} in $d=3$, or \cite{Kravec:2018qnu}),
\begin{equation} \label{eq:spectrum_excit}
	\varepsilon(n, l) = \sqrt{\frac{4 n}{d} (n + l + d - 1) + l},
\end{equation}
can be given corrections, as initiated in \cite{Orlando:2020idm}, thanks to the renormalization procedure discussed here. Similarly, the spectrum of spinning operators at large charge described in \cite{Kravec:2019djc} can be refined. We leave this for future work.

\section{Examples} \label{sec:examples}

We are now going to see this machinery in action through two examples, so let us repeat the recipe. We are interested in the conformal dimension of the lowest operator of charge $Q \gg 1$ in the theory without trap, $\Delta_Q$, which is given by the ground-state energy of the trapped system. After choosing at which order in $Q$ we want to truncate the expansion of $\Delta_Q$ (in what follows, we go up to $Q^0$), we construct the Lagrangian density by complementing Eq.~(\ref{eq:lagr_LO}) with subleading operators, Eq.~(\ref{eq:Omn_bulk}), based on the classification given in Eq.~(\ref{eq:DeltaQ_bulk}). We also include the corresponding counterterms, Eq.~(\ref{eq:Zn_edge}), and we account for the small dilaton mass deformation introduced above.

We then compute the ground-state energy density and we integrate it over the volume of the cloud, regularizing the integrals when needed. We also express the chemical potential $\mu$ as a function of the charge $Q$ to write $\Delta_Q$ as an expansion in powers of the charge and, finally, we renormalize the couplings of the counterterms so as to absorb any dependence on the regulator $\epsilon$. We do this for both $d = 2$ and $d = 3$.

\subsection{The $d=2$ case} \label{sec:ex_2D}

Typically, \acp{nrcft} in two spatial dimensions are relevant for the description of anyons \cite{Wilczek:1982wy,Nishida:2007pj}, which themselves are at the origin of the Aharonov-Bohm effect \cite{Bergman:1993kq} and whose existence has very recently been proven in the context of the fractional quantum Hall effect \cite{Bartolomei_2020,nakamura2020direct}. Based on the previous discussion, we include the subleading operators $\mathcal{O}^{(m,n)}_{bulk}$ of Eq.~(\ref{eq:Omn_bulk}) for $(m, n) \in \{ (1, 0), (0, 1), (2, 0), (1, 1) \}$, as well as the counterterms $Z^0_{edge}$ and $Z^1_{edge}$ constructed in Eq.~(\ref{eq:Zn_edge}). The Lagrangian thus reads
\[
\begin{aligned}
	\mathcal{L} & = c_\frac{3}{2} a^2 U - \frac{g}{4} a^4 - \frac{c_\frac{3}{2} m^2_\sigma}{16 f^2} \left[ (f a)^4 - 4 \log (f a) - 1 \right] \\
	& - \frac{c_\frac{1}{2}}{2} (\del_i a)^2 - \frac{c'_\frac{1}{2}}{2} Z - \frac{c''_\frac{1}{2}}{4} \frac{(\del_i a)^4}{a^4} - \frac{c_\frac{1}{6}}{4} \frac{(\del_i a)^2}{a^4} Z \\
	& + \delta(a^2) \left[ \kappa_\frac{1}{2} \frac{a^2 (\del_i a)^2}{2} + \frac{\kappa_\frac{1}{6}}{2} \left( \frac{a^2 (\del_i a)^2}{2} \right)^\frac{1}{3} Z \right],
\end{aligned}
\]
where the Wilsonian parameters of subleading operators $c_\frac{1}{2}, c'_\frac{1}{2}, c''_\frac{1}{2}$, etc. are labeled according to the order at which they enter the expansion of $\Delta_Q$ in the end, and they are normalized so as to slightly simplify the expression of the ground-state energy density $\mathcal{H}_0$. Indeed, using $\langle \chi \rangle = \mu \cdot t$, $\langle a \rangle \equiv v(z)$ and Eq.~(\ref{eq:z_stuff}), we get
\begin{equation} \label{eq:2D_H0}
\begin{aligned}
	\mathcal{H}_0 & = c_\frac{3}{2} v^2 \mu (1 - z) + \frac{\tilde g}{4} v^4 - \frac{c_\frac{3}{2} m^2_\sigma}{16 f^2} \left[ 4 \log (f v) + 1 \right] \\
	& + \frac{(1 - z) v'^2}{\mu} \left[ c_\frac{1}{2} + \frac{c_\frac{1}{6}}{v^4} \right] + c'_\frac{1}{2} + \frac{c''_\frac{1}{2} (1 - z)^2}{\mu^2} \left( \frac{v'}{v} \right)^4 \\
	& - \delta(v^2) \left[ \frac{\kappa_\frac{1}{2} (1 - z)}{\mu} (v v')^2 + \frac{\kappa_\frac{1}{6} (1 - z)^\frac{1}{3}}{\mu^\frac{1}{3}} (v v')^\frac{2}{3} \right].
\end{aligned}
\end{equation}
The \ac{vev} $v(z)$ is the solution of the \ac{eom} (cf. Eq.~(\ref{eq:eom_d}))
\begin{equation} \label{eq:2D_EoM}
	\left( \frac{v(z)}{v_{hom}} \right)^2 = z \left[ 1 + B(z,v,v',v'') \right],
\end{equation}
with $v_{hom} \equiv \sqrt{2 \mu c_\frac{3}{2} / \tilde g}$, and $B(z,v,v',v'')$ is given by
\[
\begin{aligned}
	B(z,v,v',v'') & = \frac{m_\sigma^2}{8 f^2 \mu z} \frac{1}{v^2} + \frac{c_\frac{1}{2}}{c_\frac{3}{2} \mu^2 z} \frac{(1 - z) v'' - v'}{v} \\
	& + \frac{6 c''_\frac{1}{2} (1 - z)}{c_\frac{3}{2} \mu^3 z} \frac{{v'}^2}{v^4} \frac{(1 - z) (v v'' - {v'}^2) - v v'}{v^2} \\
	& + \frac{c_\frac{1}{6}}{c_\frac{3}{2} \mu^2 z} \frac{(1 - z) (v v'' - 2 {v'}^2) - v v'}{v^6}.
\end{aligned}
\]
in the bulk. Referring to the previous section, we find that the chemical potential is related to the charge as
\[
\begin{aligned}
	\mu & = \zeta \sqrt{Q} \left[ 1 + \frac{1}{2 c_\frac{3}{2}} \frac{1}{Q^\frac{2}{3}} \left\{ \frac{c_\frac{1}{2}}{\epsilon} + \frac{3}{8} \frac{c''_\frac{1}{2} \tilde g}{c_\frac{3}{2}} \frac{1}{\epsilon^4} \right\} + \mathcal{O}\left( Q^{-1} \right) \right],
\end{aligned}
\]
where $\zeta \equiv \sqrt{\tilde g/(2 \pi c_\frac{3}{2}^2)}$. Note that the dilaton mass $m_\sigma$ modifies the expression of the chemical potential only beyond next-to-leading order. We are now in position to integrate Eq.~(\ref{eq:2D_H0}), removing the $\delta_\epsilon$-layer when necessary. Working linearly in the Wilsonian coefficients of subleading operators, we find
\begin{equation}
\begin{aligned}
	\Delta_Q^{(d=2)} & = \frac{2}{3} \zeta Q^\frac{3}{2} + \left[ \frac{c_\frac{1}{2}}{6 \zeta c_\frac{3}{2}} - \frac{\pi \zeta c_\frac{3}{2} m_\sigma^2}{8 f^2} \right] \sqrt{Q} \log Q \\
	& - \frac{k_\frac{1}{2}^{ren.}}{2 \zeta c_\frac{3}{2}} \sqrt{Q} - \left( 2 \pi^4 c_\frac{3}{2} \right)^\frac{1}{3} \zeta k_\frac{1}{6}^{ren.} \cdot Q^\frac{1}{6} - 0.29416,
\end{aligned}
\end{equation}
up to corrections scaling with negative powers of $Q$. The last term is universal and given by the Casimir energy found in \cite{Orlando:2020idm}, while the renormalized couplings are
\begin{equation}
\begin{aligned}
	\kappa_\frac{1}{2}^{ren.} & = \kappa_\frac{1}{2} + c_\frac{1}{2} \log \epsilon - \frac{\gamma}{24} \frac{c''_\frac{1}{2}}{\epsilon^3} + \mathrm{(finite)}, \\
	\kappa_\frac{1}{6}^{ren.} & = \kappa_\frac{1}{6} + \left[ \frac{c''_\frac{1}{2}}{24 \gamma^\frac{1}{3}} - \frac{\gamma^\frac{2}{3}}{8} c_\frac{1}{6} \right] \frac{1}{\epsilon^2} + \mathrm{(finite)},
\end{aligned}
\end{equation}
where $\gamma \equiv \tilde g / c_\frac{3}{2}$. Note that we have absorbed the contribution of $c'_\frac{1}{2}$, which is finite, as well as a finite correction due to $m_\sigma$ into the finite part of $\kappa_\frac{1}{2}^{ren.}$. We thus see that the effect of the small dilaton mass deformation is a mere shift in the coefficients of the $\sqrt{Q} \log Q$ and $\sqrt{Q}$-terms.

Let us mention that anyons are not invariant under parity and it would thus be interesting to extend this study to the case of parity-violating theories (cf. \cite{Kravec:2018qnu} for suggestions), in the spirit of \cite{Cuomo:2021qws} in the relativistic case.

\subsection{The $d=3$ case} \label{sec:ex_3D}

As mentioned in the introduction, the case of \acp{nrcft} in three spatial dimensions is relevant for the description of the unitary Fermi gas. In order to build the large-charge \ac{eft}, we again use the leading-order Lagrangian Eq.~(\ref{eq:lagr_LO}), to which we add Coleman's potential Eq.~(\ref{eq:Uc}), the operators given in Eq.~(\ref{eq:Omn_bulk}) with $(m, n) \in \{ (1, 0), (0, 1), (2, 0), (1, 1), (0, 2) \}$ and the edge counterterms, Eq.~(\ref{eq:Zn_edge}), constructed from $Z^0$, $Z^1$, and $Z^2$. We thus consider the following Lagrangian density:
\[
\begin{aligned}
	\mathcal{L} & = c_\frac{4}{3} a^2 U - \frac{3 \tilde g}{10} a^\frac{10}{3} + \frac{9 c_\frac{4}{3} m^2_\sigma}{100 f^2} \left[ \frac{10}{3} \log (f a) + 1 \right] \\
	& - \frac{c_\frac{2}{3}}{2} (\del_i a)^2 - \frac{c'_\frac{2}{3}}{3} a^\frac{2}{3} Z - \frac{c_\frac{5}{9}}{4} \frac{(\del_i a)^4}{a^\frac{10}{3}} - \frac{c_\frac{1}{3}}{6} \frac{(\del_i a)^2}{a^\frac{8}{3}} Z \\ 
	& - \frac{c_\frac{1}{9}}{9} \frac{Z^2}{a^2} + \delta(a^\frac{4}{3}) \left[ \kappa_\frac{5}{9} \left( \frac{a^\frac{2}{3} (\del_i a)^2}{2} \right)^\frac{7}{6} \right. \\
	& \left. + \frac{\kappa_\frac{1}{3}}{3} \left( \frac{a^\frac{2}{3} (\del_i a)^2}{2} \right)^\frac{1}{2} Z + \frac{\kappa_\frac{1}{9}}{9} \left( \frac{2}{a^\frac{2}{3} (\del_i a)^2} \right)^\frac{1}{6} Z^2 \right],
\end{aligned}
\]
where $\tilde g \equiv g \left( 1 + 3 c_\frac{4}{3} m_\sigma^2 f^\frac{4}{3} / (10 g) \right)$. The ground-state energy density then reads
\begin{equation} \label{eq:3D_H0}
\begin{aligned}
	\mathcal{H}_0 & = c_\frac{4}{3} \mu v^2 (1 - z) + \frac{3 \tilde g}{10} v^\frac{10}{3} - \frac{9 c_\frac{4}{3} m^2_\sigma}{100 f^2} \left[ \frac{10}{3} \log (f v) + 1 \right] \\
	& + \frac{{v'}^2 (1 - z)}{\mu} \left[ c_\frac{2}{3} + \frac{c_\frac{1}{3}}{v^\frac{8}{3}} \right] + c'_\frac{2}{3} v^\frac{2}{3} + \frac{c_\frac{5}{9} (1 - z)^2}{\mu^2} \frac{{v'}^4}{v^\frac{10}{3}} \\
	& + \frac{c_\frac{1}{9}}{v^2} - \delta(v^\frac{4}{3}) \left[ \frac{\kappa_\frac{5}{9} (1 - z)^\frac{7}{6}}{\mu^\frac{7}{6}} v^\frac{7}{9} {v'}^\frac{7}{3} + \frac{\kappa_\frac{1}{3} \sqrt{z}}{\sqrt{z}} v^\frac{1}{3} v' \right. \\
	& \hspace{45mm} \left. + \frac{\kappa_\frac{1}{9} \mu^\frac{1}{6}}{(1 - z)^\frac{1}{6}} \frac{1}{v^\frac{1}{9} {v'}^\frac{1}{3}} \right].
\end{aligned}
\end{equation}
Moreover, the \ac{vev} of the radial mode $v(z)$ satisfies the equation of motion Eq.~(\ref{eq:eom_d}) with $v_{hom} \equiv (2 \mu c_\frac{4}{3} / \tilde g)^\frac{3}{4}$ and
\[
\begin{aligned}
	B(z, v, v', v'') & = \frac{3 m_\sigma^2}{20 f^2 \mu z} \frac{1}{v^2} + \frac{c_\frac{2}{3}}{2 c_\frac{4}{3} \mu^2 z} \frac{2 (1 - z) v'' - 3 v'}{v} \\
	& - \frac{c'_\frac{2}{3}}{3 c_\frac{4}{3} \mu z} \frac{1}{v^\frac{4}{3}} + \frac{c_\frac{1}{9}}{c_\frac{4}{3} \mu z} \frac{1}{v^4} \\
	& + \frac{c_\frac{5}{9} (1 - z)}{c_\frac{4}{3} \mu^3 z} \frac{{v'}^2}{v^\frac{10}{3}} \frac{(1 - z) (6 v v'' - 5 {v'}^2) - 9 v v'}{v^2} \\
	& + \frac{c_\frac{1}{3}}{6 c_\frac{4}{3} \mu^2 z} \frac{(1 - z) (6 v v'' - 8 {v'}^2) - 9 v v'}{v^\frac{14}{3}}
\end{aligned}
\]
in the bulk. Consequently, the chemical potential reads
\[
\begin{aligned}
	\mu & = \zeta Q^\frac{1}{3} \left[ 1 + \frac{9}{\pi c_\frac{4}{3}} \frac{1}{Q^\frac{5}{9}} \left\{ \frac{c_\frac{2}{3}}{\sqrt{\epsilon}} + \frac{27 c_\frac{5}{9} g}{32 c_\frac{4}{3}} \frac{1}{\epsilon^\frac{7}{2}} \right\} + \mathcal{O}\left( Q^{-\frac{7}{9}} \right) \right].
\end{aligned}
\]
Proceeding as before, we finally find
\begin{equation} \label{eq:Delta_Q_3D}
\begin{aligned}
	\Delta_Q^{(d=3)} & = \frac{3}{4} \zeta Q^\frac{4}{3} + \left[ \frac{27 c_\frac{2}{3}}{8 \zeta c_\frac{4}{3}} + \frac{\pi^\frac{4}{3} \zeta c'_\frac{2}{3}}{c_\frac{4}{3}^\frac{1}{3}} \right]  \cdot Q^\frac{2}{3} \\
	& - \left[ \frac{3^{42}}{2^{15} \pi^{14} c_\frac{4}{3}^{16}} \right]^\frac{1}{18} \frac{\kappa_\frac{5}{9}^{ren.}}{\zeta} \cdot Q^\frac{5}{9} \\
	& - \frac{\sqrt{2} \pi m_\sigma^2 \zeta^\frac{3}{2} c_\frac{4}{3}}{5 f^2} \cdot \sqrt{Q} \left[ \log Q + \left( \log \frac{512 f^4}{\pi^2 \zeta^3 c_\frac{4}{3}^2} - \frac{34}{5} \right) \right] \\
	& - 3 \sqrt{2} \pi \zeta k_\frac{1}{3}^{ren.} \cdot Q^\frac{1}{3} - \left[ \frac{2^{33} \pi^{50} c_\frac{4}{3}^{16}}{3^6} \right]^\frac{1}{18} \zeta^3 \kappa_\frac{1}{9}^{ren.} \cdot Q^\frac{1}{9} \\
	& - \frac{\sqrt{2} \pi^2}{2} \left[ \frac{3^5}{16 \sqrt{\gamma}} c_\frac{5}{9} - \frac{3 \sqrt{\gamma}}{4} c_\frac{1}{3} + \frac{\gamma^\frac{3}{2}}{27} c_\frac{1}{9} \right],
\end{aligned}
\end{equation}
where $\gamma \equiv 18 \tilde g / c_\frac{4}{3}$ is a convenient parameter to express the renormalized couplings, which are given by
\begin{equation}
\begin{aligned}
	\kappa_\frac{5}{9}^{ren.} & = \kappa_\frac{5}{9} - \frac{\gamma^\frac{5}{6}}{80} \frac{c_\frac{5}{9}}{\epsilon^\frac{5}{2}} + \mathrm{(finite)} \\ 
	\kappa_\frac{1}{3}^{ren.} & = \kappa_\frac{1}{3} + \left[ \frac{135 c_\frac{5}{9}}{8 \sqrt{\gamma}} - \frac{\sqrt{2} \gamma}{18} c_\frac{1}{3} \right] \frac{1}{4 \epsilon^\frac{3}{2}} + \mathrm{(finite)} \\
	\kappa_\frac{1}{9}^{ren.} & = \kappa_\frac{1}{9} + \left[ \frac{3^5}{2^3} \frac{c_\frac{1}{3}}{\gamma^\frac{5}{6}} - \frac{3^8 \cdot 5}{2^7} \frac{c_\frac{5}{9}}{\gamma^\frac{11}{6}} - \gamma^\frac{1}{6} c_\frac{1}{9} \right] \frac{1}{\sqrt{\epsilon}} + \mathrm{(finite)}.
\end{aligned}
\end{equation}
The dilaton mass deformation $m_\sigma$ is responsible for the presence of the $\sqrt{Q} \log Q$ and $\sqrt{Q}$-terms, which is the signature of the soft breaking of conformal invariance. Moreover, there is a mixed $Q^0$-contribution that serves as a counterterm for the divergent one-loop Casimir energy, and a universal $Q^0 \log Q$-term is expected to arise as a consequence of it. This will be computed in an upcoming article \cite{Hellerman:2021qzz}.

\medskip
Among the many quantities that can be given corrections based on this construction (cf. also \cite{Son:2005rv}), let us come back to the doubly integrated density mentioned in the introduction, whose leading-order expression is given in Eq.~(\ref{eq:profile_LO}). We solve the equation of motion as in Eq.~(\ref{eq:v_NLO}), but we only account for the corrections caused $c_\frac{2}{3}$ for simplicity. We then integrate the charge density $\rho(z) = c_\frac{4}{3} v(z)^2$ over $x_1$ and $x_2$ to obtain
\begin{equation} \label{eq:linear_profile_result}
\begin{aligned}
	n(x_3) & = \frac{2 \pi g}{5} \left[ \frac{2 c_\frac{4}{3} \mu}{g} \left( 1 - \frac{x_3^2}{2 \mu} \right) \right]^\frac{5}{2} \\
	& \times \left[ 1 + \frac{45 c_\frac{2}{3}}{32 c_\frac{4}{3} \mu^2} \left\{ \frac{5}{\left( 1 - \frac{x_3^2}{2 \mu} \right)^2} - \frac{1}{\left( 1 - \frac{x_3^2}{2 \mu} \right)^3} \right\} + \ldots \right].
\end{aligned}
\end{equation}
This expression is valid for $x_3 \in [0, R_{cl} \sqrt{1 - \delta_\epsilon}]$ and can in principle improve the fitting to experimental data.

\section{Conclusion}

Via the nonrelativistic state-operator map, charged operators correspond to finite density states in a harmonic trap. In this work, we investigated the class of \acp{nrcft} whose large-charge sector is effectively described by a superfluid state in the trap, which is particularly relevant for the case of the unitary Fermi gas. Specifically, we exploited this effective description to extend all known results about the expansion of the conformal dimension of the lightest charged operator, up to quantum corrections entering at order $Q^0$, see Eq.~\eqref{eq:DeltaQ_result}, thereby uncovering a rich structure of logarithmic contributions. This is based on the recent treatment of divergences at the edge of the physical cloud of trapped particles \cite{Hellerman:2020eff}, which we reviewed and extended. We also accounted for a small dilaton mass deformation in order to explore the near-conformal regime, and illustrated the full procedure in the $d=2$ and $d=3$ cases, see Sec.~\ref{sec:examples}.

Whenever we found it appropriate, we commented on the connections with the ultracold atom literature (cf. in particular the introduction) and computed some new corrections, as in Eq.~\eqref{eq:linear_profile_result} for the doubly integrated density. This fruitful direction of research remains to be explored systematically. See also \cite{Son:2005rv,Escobedo:2009bh}.

Let us finally mention the seminal works \cite{Son:2008ye,Balasubramanian:2008dm} paving the way toward a gravity/\ac{nrcft} correspondence (see also \cite{PhysRevD.85.106001} for a recent proposal). It would be fascinating to understand whether the large-charge sector of certain \acp{nrcft} can be described in a dual picture and how this would relate to the effective construction used here.

\subsection*{Acknowledgments}

Enlightening discussions with Simeon Hellerman, Domenico Orlando, Susanne Reffert and Ian Swanson are gratefully acknowledged. This work is supported by the Swiss National Science Foundation under grant No. 200021 192137.

\bibliographystyle{utphys}
\bibliography{References}

\end{document}